\documentclass[%
 aip,
 jcp,
 amsmath,amssymb,
 reprint,%
floatfix
]{revtex4-2}

\usepackage{graphicx}
\usepackage{dcolumn}
\usepackage{bm}
\usepackage[T1]{fontenc}
\usepackage{mathptmx}
\usepackage{etoolbox}
\usepackage[inline]{enumitem}
\usepackage{chemformula}

\makeatletter
\def\@email#1#2{%
 \endgroup
 \patchcmd{\titleblock@produce}
  {\frontmatter@RRAPformat}
  {\frontmatter@RRAPformat{\produce@RRAP{*#1\href{mailto:#2}{#2}}}\frontmatter@RRAPformat}
  {}{}
  
}%
\makeatother
\newcommand{\taumd}{\tau}
\begin{document}

\title{Relationship between Structure and Dynamics  of an Icosahedral Quasicrystal using Unsupervised Machine Learning}
\author{Edwin A. Bedolla-Montiel}
\email{e.a.bedollamontiel@uu.nl}
\affiliation{
  Soft Condensed Matter \& Biophysics, Debye Institute for Nanomaterials Science, Utrecht University, Princetonplein 1, 3584 CC Utrecht, the Netherlands
}%
\author{Susana Marín-Aguilar}%
\email{susana.marinaguilar@uniroma1.it}
\affiliation{
  Department of Physics, Sapienza University of Rome, Rome 00185, Italy
}%
\author{Marjolein Dijkstra}
\email{m.dijkstra@uu.nl}
\affiliation{
  Soft Condensed Matter \& Biophysics, Debye Institute for Nanomaterials Science, Utrecht University, Princetonplein 1, 3584 CC Utrecht, the Netherlands
}%


\begin{abstract}
We present a comprehensive study of the structure, formation, and dynamics of a one-component model system that self-assembles into an icosahedral quasicrystal (IQC). Using molecular dynamics simulations combined with unsupervised machine learning techniques, we identify and characterize the unique structural motifs of IQCs, including icosahedral and dodecahedral arrangements, and quantify the evolution of local environments during the IQC formation process. Our analysis reveals that the formation of the IQC is driven by the emergence of distinct local clusters that serve as precursors to the fully developed quasicrystalline phase. 
Additionally, we examine the dynamics of the system  across a range of temperatures, identifying transitions from vibrationally restricted motion to activated diffusion, and uncovering signatures of dynamic heterogeneity inherent to the quasicrystalline state.
To directly connect structure and dynamics, we use a machine-learning-based order parameter to quantify the presence of distinct local environments across temperatures. We find that regions with high structural order, as captured by specific machine-learned classes, correlate with suppressed self-diffusion and minimal dynamical heterogeneity, consistent with phason-like motion within the IQC. In contrast, regions with lower structural order exhibit enhanced collective motion and increased dynamical heterogeneity. These results establish a quantitative framework for understanding the coupling between structural organization and dynamical processes in quasicrystals, providing new insights into the mechanisms governing IQC stability and dynamics.

\end{abstract}

\maketitle

\section{\label{sec:intro}Introduction}

Icosahedral quasicrystals (IQCs) are a unique class of solids that exhibit long-range icosahedral order without translational periodicity.~\cite{janot1997quasicrystals} First discovered in metallic alloys,~\cite{shechtmanMetallicPhaseLongRange1984,levineQuasicrystalsNewClass1984} IQCs challenge the conventional classification of materials as either crystalline or amorphous. Unlike periodic crystals, which possess strict translational symmetry, IQCs display an aperiodic yet highly ordered arrangement governed by icosahedral symmetry.~\cite{ishimasaNewOrderedState1985,angellAmorphousStateEquivalent2000} These properties position IQCs as an intermediate form of matter, exhibiting a higher degree of order than  solids while lacking the periodicity of crystals.  This distinctive structural organization endows IQCs with exceptional  physical properties, including photonic wave confinement~\cite{jeonIntrinsicPhotonicWave2017} and light localization,~\cite{sinelnikExperimentalObservationIntrinsic2020} making IQCs  promising candidates for applications in optics, catalysis, and advanced material design.~\cite{yablonovitchInhibitedSpontaneousEmission1987,manExperimentalMeasurementPhotonic2005}

Research on IQCs has predominantly focused on their thermodynamic and structural properties, with extensive efforts being made to understand their stability and explore their potential applications~\cite{mikheevaThermodynamicKineticProperties2000,tsaiDiscoveryStableIcosahedral2013}. It is now well established that IQCs can be stabilized by specific atomic-scale interactions as well as in model systems of colloidal and nanoparticle self-assembly~\cite{tsaiIcosahedralClustersIcosaheral2008,dshemuchadseMovingConstraintsChemistry2021,noyaHowDesignIcosahedral2021,pinto2024automating, noya2025one}. While much is known about their structural organization,  their dynamical properties have received less attention, despite their critical role in phenomena such as self-assembly, defect migration, and structural stabilization.
Theoretical approaches have attempted to describe quasicrystal dynamics using phase field crystal models, which capture density wave interactions and provide insight into equilibrium stability~\cite{subramanianThreeDimensionalIcosahedralPhase2016}. However, these models do not fully resolve the microscopic mechanisms driving particle motion, nor do they account for the dynamic heterogeneity observed in IQCs at finite temperatures~\cite{zhao2025atomistic}. Experimental investigations into the mechanical response of icosahedral Al-Pd-Mn quasicrystals have revealed complex elastodynamic behavior, suggesting that phason excitations play a key role in structural rearrangements~\cite{qiaoDynamicResponseIcosahedral2017,han2021formation}. Yet, how these phasonic modes couple to self-diffusion and how they influence the long-term stability of IQCs remain open questions. Understanding these aspects is crucial for predicting how IQCs form, evolve, and adapt to changing conditions, as their unique aperiodic order implies fundamentally different diffusion and relaxation mechanisms compared to periodic crystals.

A major theoretical insight into IQC dynamics was proposed by Kalugin and Katz~\cite{kaluginMechanismSelfDiffusionQuasiCrystals1993}, who demonstrated that self-diffusion in IQCs does not follow conventional vacancy-mediated mechanisms but instead proceeds through an energy-activated process. Their work, later validated experimentally~\cite{bluherFirstLowTemperatureRadiotracer1998,brandDynamicsIcosahedralQuasicrystal2001}, revealed that IQCs exhibit deviations from Arrhenius-type diffusion, where particle motion arises from a combination of phason-induced rearrangements and collective displacements~\cite{engelDynamicsParticleFlips2010}. While computational studies have further extended these ideas in two-dimensional (2D) quasicrystals, where diffusion is mediated by both individual particle jumps and correlated defect movements~\cite{zhao2025atomistic}, a detailed understanding of these mechanisms in three-dimensional IQCs remains incomplete.

Another fundamental challenge in IQC research lies in the absence of a robust order parameter capable of capturing their structural evolution. In periodic crystals, translational order can be quantified straightforwardly, but the complex, aperiodic arrangement of IQCs presents a greater challenge. Conventional structural characterization methods, such as diffraction and radial distribution functions, provide valuable but incomplete descriptions of local order. Machine learning techniques have emerged as a promising alternative, enabling the identification of subtle local structural motifs that are difficult to detect using traditional methods. Unsupervised learning algorithms, in particular, have proven effective in classifying structural environments in glassy systems~\cite{boattiniAutonomouslyRevealingHidden2020a,coslovichDimensionalityReductionLocal2022b,ciarella2023dynamics} and quasicrystals~\cite{spellingsMachineLearningCrystal2018}, offering a data-driven approach to defining an IQC-specific order parameter.
One major unknown is how to quantitatively characterize the diverse local environments in QCs and to relate these structural motifs to dynamical processes such as diffusion, phason activity, and dynamic heterogeneity. While recent progress has been made using molecular simulations and spectral analyses to study these phenomena, most approaches rely on predefined order parameters or symmetry assumptions, limiting their ability to capture the full complexity of quasicrystalline order.

In this work, we address this gap by asking: \emph{To what extent does the local structure of a quasicrystal determine its dynamical behavior?} We explore this question using molecular dynamics simulations of a three-dimensional IQC model that exhibits a rich spectrum of structural and dynamical behavior~\cite{engel_computational_2015}.
By combining structural analysis with unsupervised machine learning, we establish a novel framework for characterizing quasicrystal structures. This approach enables us to systematically distinguish between IQCs, fluids, crystals, and amorphous phases present in the system's phase diagram. In addition, it allows us to identify local structures that act as precursors during IQC formation, thereby enabling a full characterization of the transformation pathways that lead to quasicrystalline order. 
We analyze the self-diffusion and dynamic heterogeneity of IQCs, and find that the two distinct diffusion regimes observed in single-particle trajectories are closely associated with dynamical  heterogeneity.
Moreover, we establish a quantitative connection between structure and dynamics by analyzing how local structural environments, identified via an unsupervised machine learning framework, correlate with particle mobility. We show that specific classes of local environments correspond to distinct dynamical behaviors, revealing how structural order governs the extent of dynamical heterogeneity within the quasicrystal.

The paper is organized as follows. Section~\ref{sec:methods} describes the simulation model and methodology. Section~\ref{sec:phases} presents the phase behavior of the IQC system and its structural properties, followed by Section~\ref{sec:uml}, where we introduce a machine learning approach to classify local environments within the IQC. Section~\ref{sec:nucleation} explores the formation process of the quasicrystal, while Section~\ref{sec:dynamics} examines its dynamical properties, including self-diffusion and dynamical heterogeneity.
In Section~\ref{sec:structure-dynamics}, we establish a direct connection between structure and dynamics by analyzing how machine-learned structural classes correlate with particle mobility and dynamical behavior. Finally, Section~\ref{sec:conclusions} discusses our findings and outlines future research directions.

\section{\label{sec:methods}Model and methods}
\subsection{Model}
We consider a three-dimensional one-component system of particles interacting via an isotropic pair potential characterized by three oscillations~\cite{engel_computational_2015}.
This potential introduces two distinct length scales, facilitating the formation of an icosahedral quasicrystal. Its functional form is inspired by Friedel oscillations, analogous to the Hume-Rothery mechanism, which effectively mimics the atomic interactions of metallic systems that form quasicrystalline structures at zero temperature~\cite{mihalkovivc2012empirical}.
The potential combines short-range repulsion and damped oscillations with wavenumber \( k \) and a phase shift \(\phi\), which generate attractive interactions at specific interparticle distances \( r \). The interaction potential is given by  
\begin{eqnarray}
V(r) / \epsilon &=& \nonumber \\
& &\hspace{-10mm}
\begin{cases}
 {\left(\frac{\sigma}{r}\right)}^{15} + {\left(\frac{\sigma}{r}\right)}^{3} \cos{\left( k  \left(r - 1.25\sigma \right) - \phi \right)} &r < r_{c} \\
 \hspace{2.5cm} 0 &r \geq r_{c} 
\end{cases}
\, ,
    \label{eq:potential}
\end{eqnarray}
where \( \sigma \) denotes the particle diameter, \( \epsilon \) is a measure for the interaction strength, and the cutoff radius is set to \(r_{c} = 2.82 \sigma\), beyond which the potential is zero. The potential is truncated after the third maximum, and smoothly shifted to zero.
\begin{figure}
    \includegraphics{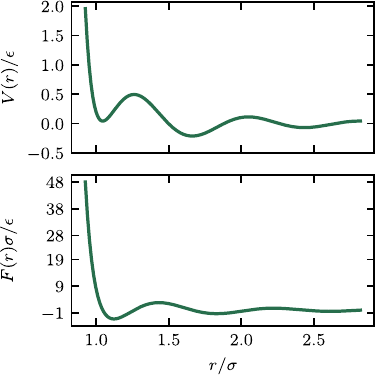}
    \caption{Interaction potential \(V(r)/\epsilon\) as a function of distance $r$, as defined in Eq.~\ref{eq:potential} for the parameters \( k \sigma = 8.21\) and \(\phi = 0.46\) and a cutoff radius of \(r_{c} = 2.82 \sigma\). The bottom panel shows the force \(F(r)\sigma/\epsilon\) as a function of distance $r$.}
    \label{fig:inter-pot}
\end{figure}
This ensures that both the potential and its first derivative, namely the force, remain continuous across the interaction range, as shown in Fig.~\ref{fig:inter-pot}.
We set the parameter values \( k  \sigma = 8.21\) and \(\phi = 0.46\), which promote the spontaneous formation of an icosahedral quasicrystal from a dilute gas phase at zero pressure~\cite{engel_computational_2015}.

\subsection{Simulation details}
To study the structural, thermodynamic, and dynamical properties of icosahedral quasicrytals, we perform \(NVT\)\ and \(NPT\)\ molecular dynamics simulations using the LAMMPS software~\cite{THOMPSON2022108171}. All particle visualizations are generated with OVITO~\cite{ovito}.
Each simulation consists of \(N = 4096\) particles with mass \(m\) and diameter \(\sigma\). 
Unless otherwise specified, periodic boundary conditions are applied in all directions. 

To keep the temperature constant, we use the Bussi-Donadio-Parrinello
thermostat~\cite{bussiCanonicalSamplingVelocity2007} with a damping constant of $ 0.5 \, \taumd $, where we define $\taumd = \sqrt{m \sigma^{2} / \epsilon}$ as the unit of time. To maintain constant pressure, we use the equations of motion by~\textcite{shinodaRapidEstimationElastic2004}, 
as implemented in LAMMPS~\cite{THOMPSON2022108171}, with a barostat constant of $ 5 \, \taumd $. The time step for all simulations is set to $\delta \tau = 0.005 \, \taumd$.

We employ a cooling-compression protocol to produce low-energy icosahedral quasicrystals with minimal defects. Starting from a low-density fluid at temperature \(k_B T / \epsilon = 1\)  and pressure \(P \sigma^3 / \epsilon = 0.01\), the system is gradually cooled and compressed over \(10^9 \taumd\), reaching  \(k_B T / \epsilon = 0.1\) and \(P \sigma^3 / \epsilon = 0.1\). During this process, the density of the fluid increases and the system undergoes a phase transition into a quasicrystalline state.
Once the quasicrystal is formed, the system is equilibrated in the \(NPT\)\ ensemble for \(10^7 \taumd\).
Throughout the cooling and compression, the energy per particle decreases, stabilizing the quasicrystalline phase.
Additional details are provided in Appendix~\ref{ap:nucleation}.
The resulting low-energy icosahedral quasicrystal at temperature \(k_BT / \epsilon = 0.1\) and pressure \(P \sigma^3/\epsilon = 0.1\) is employed as the initial configuration for subsequent simulations to investigate the structural, thermodynamic, and dynamical properties.
We gradually increase the temperature in the \(NPT\) ensemble over \( 5 \times 10^7 \taumd\)  to reach the target temperature while keeping the pressure fixed at \( P \sigma^3 / \epsilon = 0.1 \).
After this heating process,   an equilibration step is followed  for an additional \( 5 \times 10^7 \taumd \) in the 
\(NVT\) ensemble.
This procedure is repeated for temperatures within the range of \(k_{B} T / \epsilon \in [0.1, 0.3]\). 
At temperatures above this range, the system remains in a fluid phase with no signs of quasicrystal formation.
Finally, for each temperature, we collect data  for  dynamical quantities over at least \(5 \times 10^8 \taumd\) in the \(NVT\) ensemble.

\subsection{\label{sec:order}Structural Analysis}
To distinguish and characterize the global structure of the various phases present in  this system, we calculate the radial distribution function (RDF) defined as
\begin{equation}
    g(r) = \frac{1}{\rho} \left\langle \frac{1}{N} \sum_{i=1}^{N}
    \sum_{i \neq j}^{N} \delta(r - \lvert {\bf r}_{i} - {\bf r}_{j} \rvert)  \right\rangle ,
    \label{eq:rdf}
\end{equation}
where $\mathbf{r}_{i(j)}$ is the position of particle $i(j)$ and $\rho=N/V$ is the number density of the system.

In addition, we characterize the symmetry of each phase by computing its diffraction pattern.
In particular, to obtain the diffraction pattern, we first align the simulation box along a symmetry axis and then project the particle positions onto a plane after the rotation. These projected positions are subsequently used to compute the structure factor, which is defined as  
\begin{equation}
    S({\bf q}) = \frac{1}{N} \left\langle {\left\lvert \sum_{j=1}^{N} 
    \exp{\left(-i \, {\bf q} \cdot {\bf r}_{j}(t) \right)} \right\rvert}^{2} \right\rangle ,
    \label{eq:sq}
\end{equation}
where \({\bf q}\) is the wave vector, and the angular brackets denote an average  over configurations taken at different times within the same simulation.

Finally, we calculate the bond orientational order parameters (BOPs), to generate local descriptors that can be used as inputs for the unsupervised machine learning algorithms. These parameters are based on the framework  introduced by Steinhardt \emph{et al.}~\cite{steinhardtBondorientationalOrderLiquids1983}, and are defined as
\begin{equation}
    q_{lm} (i) = \frac{1}{N_{b} (i)} \sum_{j=1}^{N_{b}(i)} Y_{l}^{m} ({\bf r}_{ij}) 
    \, ,
    \label{eq:qlm}
\end{equation}
where \(Y_{l}^{m} (\bm{r}_{ij})\) are the spherical harmonics with total angular momentum \(l\) and  projection \(-l \leq m \leq l\), $\mathbf{r}_{ij}$ is the vector from particle $i$ to particle $j$, and $N_{b}(i)$ represents the number of nearest neighbors of particle $i$. Here, we fix $N_b$ to the  $12$ nearest neighbors. The rotationally invariant BOPs are then defined as,
\begin{equation}
    q_{l} (i) = \sqrt{\frac{4 \pi}{2l + 1} \sum_{m=-l}^{l} {\left\lvert q_{lm} (i) \right\rvert}^{2}} 
    \, ,
    \label{eq:ql}
\end{equation}
To incorporate information from the second neighbor shell, we compute the averaged BOPs~\cite{boattiniAutonomouslyRevealingHidden2020a},

\begin{equation}
    \bar{q}_{l}(i) = \frac{1}{N_{b}(i) + 1} \left[ q_{l}(i) + \sum_{k}^{N_b(i)} q_{l}(k) \right] \, .
    \label{eq:ql_average}
\end{equation}

\section{\label{sec:phases}Phase behavior and Structure}
We begin our investigation by examining the phase behavior of the system. To this end, we compute the equations of state at two different temperatures.  Starting from a low-density fluid at \(k_{B} T / \epsilon = 1\) and \(P\sigma^3/\epsilon=0.01\),  we gradually cool and compress the system until the desired temperature and pressure are achieved  using the cooling and compression protocol described in Appendix \ref{ap:nucleation}. Once the desired  state point is achieved, we characterize the structure of the resulting phase by computing the radial distribution function (RDF)  and its corresponding diffraction pattern.

\begin{figure}
    \includegraphics{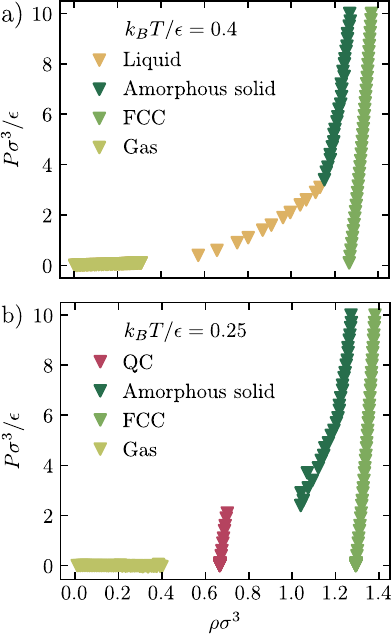}
    \caption[eos]{Equations of state at temperature
    \begin{enumerate*}[label={\alph*})]
        \item  \(k_{B} T / \epsilon = 0.4\): The system undergoes a gas-to-liquid transition at low densities, followed by the formation of an amorphous solid, and eventually crystallizes into an FCC at high densities\label{fig:no-qc-eos}, and
        \item  temperature \(k_{B} T / \epsilon = 0.25\): An icosahedral quasicrystal is observed within a narrow density range between the gas phase and the amorphous solid. At higher densities, the system crystallizes into an FCC crystal.\label{fig:qc-eos}
    \end{enumerate*}}
    \label{fig:eos}
\end{figure}

\begin{figure*}
    \includegraphics{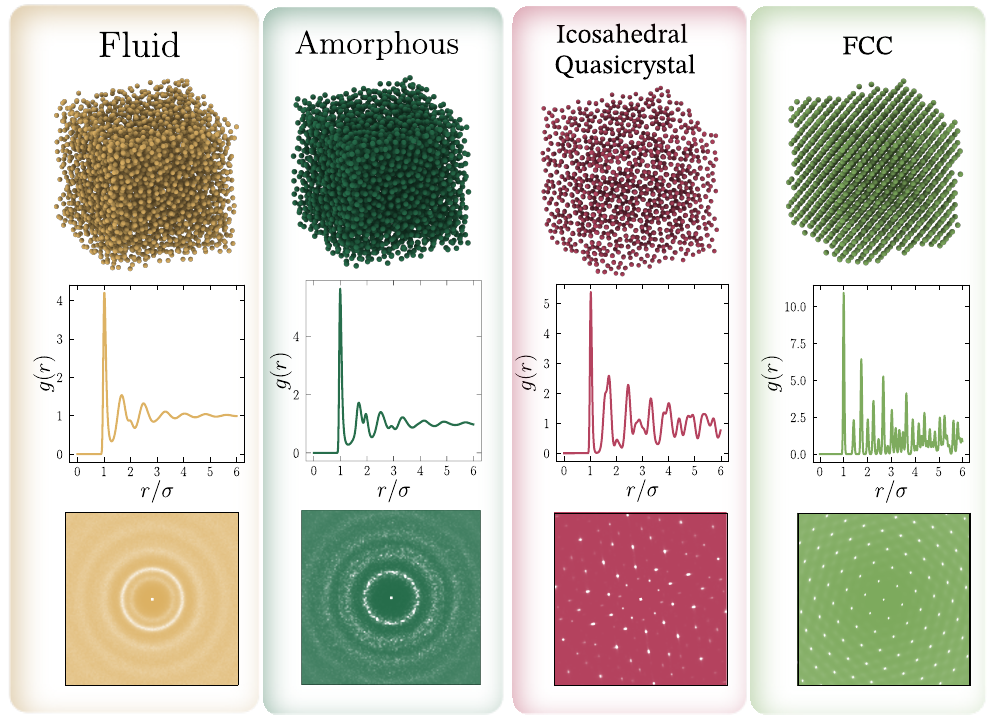}
    \caption{Phases observed for  a system of particles interacting via the potential described in Eq.~\ref{eq:potential}, along with their corresponding  radial distribution functions and  diffraction patterns. The liquid phase, shown  in the first column, is observed at \(k_{B} T / \epsilon = 0.45\) and \(P \sigma^{3} / \epsilon = 0.1 .\) The amorphous solid is found at \(k_{B} T / \epsilon = 0.45\) and \(P \sigma^{3} / \epsilon = 10 .\) The icosahedral quasicrystal, oriented along its five-fold symmetry axis (also used for computing the diffraction pattern), is obtained at  \(k_{B} T / \epsilon = 0.25\) and \(P \sigma^{3} / \epsilon = 0.1 .\) Finally, the FCC crystal, shown in the last column, is observed at  \(k_{B} T / \epsilon = 0.25\) and \(P \sigma^{3} / \epsilon = 10 .\)}
    \label{fig:structure}
\end{figure*}

At a relatively high temperature of \(k_{B} T / \epsilon = 0.4\), the system exhibits a rather simple phase behavior. 
We observe a gas phase at low pressure, which transitions into a fluid phase upon increasing the pressure, as shown in Fig.~\ref{fig:eos}a.
Upon further compression, the system  becomes dynamically arrested, forming an amorphous solid. 
We distinguish the amorphous phase from the fluid phase by analyzing their structural signatures. Although these two phases exhibit similar structural features--specifically, a lack of long-range order in the RDF and ring-like patterns in the diffraction patterns characteristic of isotropic phases, shown in the first two columns of Fig.~\ref{fig:structure}--key differences emerge. In the amorphous solid, the RDF exhibits a splitting of the second peak, indicating the presence of icosahedral-like local environments. This is  consistent with the shape of the interaction potential and similar to structures observed in certain glass-forming systems\cite{royall2015role,marin2019slowing}. These icosahedral-like motifs act as locally favored structures that are incompatible with the long-range positional order of crystalline structures, giving rise to the observed amorphous solid-like behavior. Additionally, the diffraction pattern of the amorphous solid displays speckles within the rings, indicating the onset of local particle ordering.
At  higher densities, we find  a stable face-centered-cubic (FCC) crystal phase, which reverts to a gas phase as the pressure decreases, as shown in Fig.~\ref{fig:eos}a. The long-range order of the FCC crystal is evident in the periodic peaks of the RDF and the six-fold symmetric diffraction pattern, as shown in the last column of Fig.~\ref{fig:structure}.
Across this temperature range, we observe no evidence of quasicrystalline order.
At a lower temperature of \(k_{B} T / \epsilon = 0.25\), the system exhibits  different behavior. Upon compression, the gas phase transitions into an IQC, which remains stable within a narrow density range, as shown in the equation of state  in Fig.~\ref{fig:eos}b. This phase is  structurally  distinct from the phases discussed previously. We note, however, that the IQC formed directly from compression of the gas phase at this temperature tends to exhibit defects and features only partial icosahedral order. To reduce the amount of  defects, we employ the  cooling and compression protocol  described above. Starting from a low-density, high-temperature configuration, we gradually cool and compress the system   to 
\(k_{B} T / \epsilon = 0.1\) and \(P\sigma^3/\epsilon=0.1\),  respectively, facilitating the formation of a low-energy IQC with minimal defects. We then thermalize this configuration to  \(k_{B}T/\epsilon=0.25\). The resulting low-energy IQC remains stable even at pressures approaching zero.

 The IQC displays long-range aperiodic order with icosahedral symmetry, as shown in the third column of Fig.~\ref{fig:structure}. A representative configuration, oriented along the five-fold symmetry axis, reveals the characteristic icosahedral motifs for a system at $k_BT/\epsilon = 0.25$ and $P\sigma^3/\epsilon=0.1$. This structural order is also reflected in the RDF, which displays sharp peaks extending over long distances. In particular, the peaks corresponding to the second shell show signatures of icosahedral and dodecahedral ordering, while  peaks at larger distances indicate  the presence of more complex structures, such as truncated icosahedra and tricontahedra~\cite{engel_computational_2015}.
 Finally, the diffraction pattern along the five-fold symmetry axis exhibits distinctive peaks corresponding to the ten-fold symmetry, consistent with previous experimental and simulation observations of IQCs~\cite{shechtmanMetallicPhaseLongRange1984,engel_computational_2015,cao2025phonon}.

Upon further compression, the quasicrystal loses most of its icosahedral order
and transforms into an amorphous solid. At even higher densities, similar to the behavior at higher temperatures, the system transitions into an FCC crystal phase. 

\section{\label{sec:uml}Unsupervised Learning of Local Order in Icosahedral Quasicrystals}
As we have seen, a defining characteristic of quasicrystals is the absence of long-range periodic order, accompanied by a wide variety of local structural environments. This makes the characterization of IQCs particularly challenging, as it requires the identification of the specific symmetry axes associated with icosahedral order. Consequently, there is no unique order parameter that can simultaneously distinguish the IQC from other phases and capture the diversity of its local environments.
To address these challenges, we adopt an unsupervised machine learning (UML) approach based on averaged bond orientational order parameters (BOPs)\cite{steinhardtBondorientationalOrderLiquids1983}. This method has been successfully applied to the classification of crystal structures\cite{boattini2019unsupervised} and to identifying local structures in glass-forming liquids~\cite{boattiniAutonomouslyRevealingHidden2020a}.

We first compute the BOPs defined in Eq.~\eqref{eq:ql_average} for each particle $i$, using spherical harmonics of order \(l \in [2, 12]\), resulting in a feature vector $\mathbf{Y}(i)$ for each particle.  To construct the dataset, we sample configurations from each phase and collect the BOP vectors for all $N=4096$ particles, yielding a high-dimensional dataset of size \( (N, 11) \).

To classify structural phases and identify distinct local environments in the quasicrystal, we employ an unsupervised machine learning (UML) framework based on bond orientational order parameters. Specifically, we use the Uniform Manifold Approximation and Projection (UMAP)~\cite{mcinnes2020umapuniformmanifoldapproximation} to reduce the dataset from \(11\) features to \(2\), effectively creating a two-dimensional dataset, enabling visual identification and separation of structural clusters.
Following the dimensionality reduction, we apply a Gaussian mixture model~\cite{hastie2017elements,scikit-learn} to cluster the projected data. This probabilistic approach captures the underlying phase structure by modeling the distribution as a combination of multiple Gaussian components. The number of components is chosen to match the number of expected phases in the system, which is four in our case. Details on the parameter choices for UMAP and the clustering procedure are provided in Appendix~\ref{appendix:umap}.

In Fig.~\ref{fig:phase-classification}, we show  the projection and clustering in the low-dimensional space using data from the fluid,  amorphous solid,  IQC, and  FCC phases.
Notably, particles from the same phase cluster into well-defined, distinct regions, indicating that the BOPs provide sufficient information to differentiate between the local environments of each phase. More importantly, this method also captures subtle distinctions between phases that share similar local features. This is evident in the case of the fluid and the amorphous solid, which, as  shown in the previous section, exhibit nearly identical structural signatures. 

Now that we have an order parameter capable of distinguishing between the different phases in the system, we aim to extend this  approach to identify distinct local environments within the IQC.
In this way, we  can  effectively resolve  the various local environments that emerge during  quasicrystal formation, providing valuable insights into the kinetic pathways underlying its structural transformation.  

\begin{figure}
     \includegraphics[width=0.95\linewidth]{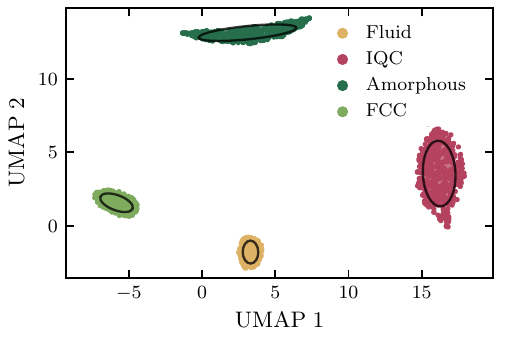}
    \caption{Dimensionality reduction and clustering of the icosahedral quasicrystal (IQC), the fluid, the amorphous solid, and the face-centered cubic (FCC) phases. The dimensions represent the embedded space obtained  from the dimensionality reduction step of the machine-learning framework.}
    \label{fig:phase-classification}
\end{figure}

\section{\label{sec:nucleation} Structural Evolution During Icosahedral Quasicrystal Formation}
\begin{figure*}
     \includegraphics[width=\linewidth]{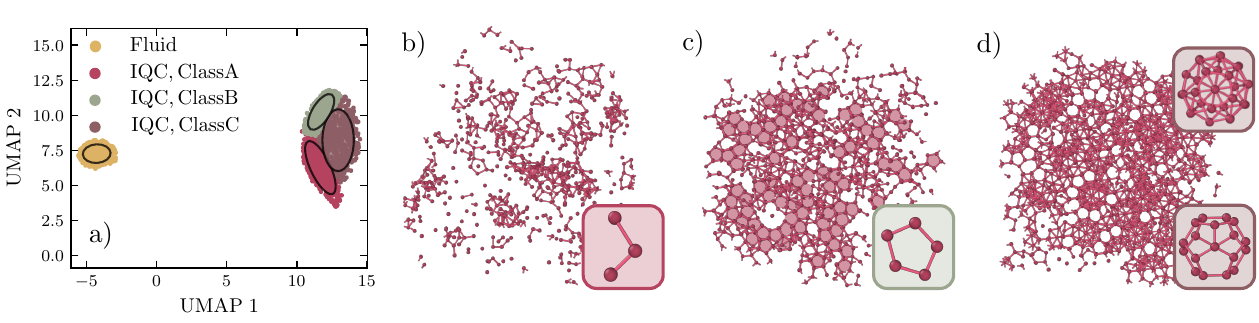}
    \caption[qc-env]{Classification of phases and quasicrystal environments using unsupervised machine learning. 
    \begin{enumerate*}[label={\alph*})]
     \item Dimensionality reduction and clustering of the icosahedral quasicrystal (IQC) and the fluid phases.  The three distinct classes A, B, and C represent different local environments identified by the machine-learning framework.
        \item Snapshot with particles classified as class A, \item B, and \item C. The insets show recurrent motifs found in each class. In particular, class A presents low coordination motifs, precursors of more complex structures. Particles in class B are mostly pentagons, which are colored for an enhanced view of the Penrose tiling that the particles appear to form. Class C corresponds to high coordination structures, that are part of larger clusters of icosahedra and dodecahedra.
    \end{enumerate*}
    }
    \label{fig:qc-env}
\end{figure*}
\begin{figure}[!b]
    \includegraphics{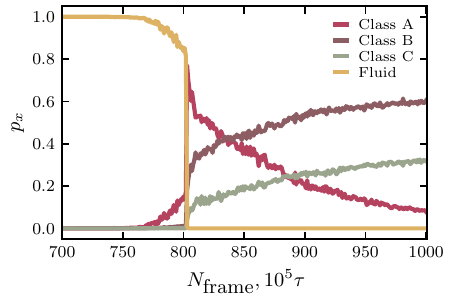}
    \caption[qc-class]{
        Time evolution of the fraction of particles belonging to a specific class, \(p_{x}\), where \(x\) denotes either fluid, class A, B or C. The sharp transition indicates the spontaneous transformation of the fluid into the quasicrystal.}
    \label{fig:nucleation}
\end{figure}
To gain  deeper insight into  the distinct local environments within the IQC, we train our UML framework using configurations from the quasicrystal at different temperatures, \(k_{B} T / \epsilon = 0.1, 0.22, 0.3\), along with a sample from the fluid phase at \(k_{B} T / \epsilon = 0.4\). This selection allows us to identify the various  local environments within the IQC and to track the IQC formation process.

Fig.~\ref{fig:qc-env}a shows the projection of this new dataset onto the low-dimensional space. The clustering procedure reveals three distinct classes within the quasicrystal, shown in Fig.~\ref{fig:qc-env}b-\ref{fig:qc-env}d, in addition to the 
fluid phase. We use this classification to track the structural evolution of a system undergoing the cooling-compression protocol that leads to quasicrystal formation. Throughout the simulation, particles are tagged according to their UML classification. To quantify the structural  evolution, we define a simple order parameter,  
 \(p_{x}\), as the fraction of particles classified as fluid, IQC class A, class B or class C, relative to the total number of particles, where $x$ denotes the  respective phase. The time evolution of \(p_{x}\) is shown in Fig.~\ref{fig:nucleation}.

 Initially, the system is in a low-density disordered state at \(k_BT/\epsilon=1\) and \(P\sigma^3/\epsilon=0.01\), where all particles are classified as fluid, so \(p_{\text{fluid}} = 1\). As the system is gradually cooled and compressed, the IQC class A begins to appear around \(t\approx 8 \times 10^{7} \taumd\), indicating the onset of local ordering. To characterize these structures, we compute their coordination number by counting the number of neighbors within a cutoff radius of \(r_{c}=1.1 \sigma\). Class A particles have an average coordination number of \(3\). This class corresponds to  precursors of pentagonal motifs, as shown in Fig.~\ref{fig:qc-env}b, which serve as  building blocks for icosahedra and dodecahedra in the quasicrystal.

Finally, we observe a sharp transition at \(t \approx 10^{8} \taumd\), where the quasicrystal spontaneously forms from the fluid phase. At this point, \(p_{\text{fluid}}\) drops to zero, and IQC classes B and C emerge, as shown in Fig.~\ref{fig:nucleation}. Although the quasicrystal  forms at this transition point, its structure continues to evolve, as seen in the ongoing changes in \(p_{A}\), \(p_{B}\) and \(p_{C}\). Over time, the fraction of class A diminishes, indicating that its structural information becomes redundant as the more  developed environments of classes B and C dominate.
Classes B and C are associated with higher coordination numbers of \(4\) and \(5\), respectively. Class B primarily identifies  networks of strongly correlated pentagons, corresponding to  the Penrose tiling forming along  the five-fold symmetry axis characteristic of the IQC~\cite{engel_computational_2015}. In Fig.~\ref{fig:qc-env}c,  these pentagons  are highlighted in  darker color. The ability of the UML framework to distinguish such tilings indicates that it captures highly symmetric, non-periodic structural motifs specific to the IQC. 
Class C, in contrast, captures highly coordinated and complex environments, such as icosahedral and dodecahedral clusters, as shown in Fig.~\ref{fig:qc-env}d. Importantly,  classes B and C are absent in the fluid phase,  implying that high coordination  environments like icosahedra are not precursors to quasicrystal formation but instead  emerge after pentagonal motifs become prevalent. These pentagons then serve as the foundation for the more complex local order observed in the IQC.

This interpretation aligns with previous studies on IQC nucleation in both experimental systems, such as the alloy \ch{Al74Pd20Mn6}~\cite{senabulyaGrowthInteractionsIcosahedral2019} and in one-component IQC simulations~\cite{liangMolecularLevelInsightsNucleation2022}. These findings support the notion that IQC formation proceeds via a two-step process,  driven by the assembly and coalescence of intermediate-sized ordered domains that form upon physical contact~\cite{han2021formation}.

\section{\label{sec:dynamics} Dynamical behavior}

Having characterized the structural environments and the kinetic pathway of the IQC transformation, we now turn to the dynamical behavior of the system. Due to their aperiodic structure, quasicrystals exhibit inherently complex dynamics. At  low temperatures, particle motion is governed by phason dynamics~\cite{de2012phonons}, while at higher temperatures, the dynamics becomes more intricate, exhibiting glassy-like features~\cite{zhaoQuasicrystalsIntermediateForm2024}.

Here, we aim to quantify these mechanisms by examining the dynamical behavior of the IQC across temperatures ranging from $k_BT/\epsilon=0.1$, where the IQC is stable, up to $k_BT/\epsilon=0.3$, where the IQC structure begins to degrade. 

\subsection{Diffusion Mechanisms}
We first examine the global dynamics of the system by analyzing the mean square displacement (MSD) of the IQC, defined as
\begin{equation}
    \left\langle r^{2}(t) \right\rangle = \frac{1}{N} \left\langle \sum_{i=1}^{N} {\left\lvert {\bf r}_{i} - {\bf r}_{i}(0) \right\rvert}^{2} \right\rangle \, .
    \label{eq:msd}
\end{equation}

\begin{figure}
     \includegraphics{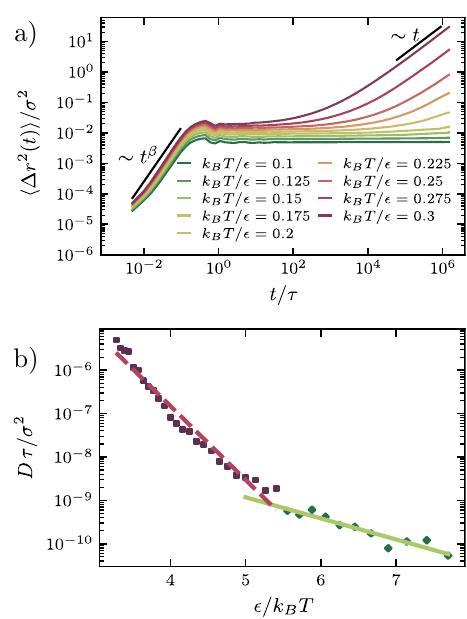}
    \caption[msd]{\begin{enumerate*}[label=\alph*)]
        \item Mean square displacement (MSD) of particles in the icosahedral quasicrystal at different  temperatures \(k_{B} T / \epsilon\). In the short-time regime, the MSD exponent \(\beta\) remains consistently below  two across all temperatures.\label{fig:msd}
        \item Long-time self-diffusion coefficient as a function of  inverse temperature. The lines are non-linear fits to the data, with different symbols indicating distinct fitting coefficients.\label{fig:diff}
        \end{enumerate*}}
    \label{fig:msd-diffusion}
\end{figure}

Fig.~\ref{fig:msd-diffusion}a shows  $\left\langle r^{2}(t) \right\rangle$ at  different temperatures. 
In the short-time regime, where  ballistic motion is expected, particles exhibit super-diffusive behavior, which we  characterize by fitting  a power-law relation  $\langle r^{2}(t)\rangle \sim t^{\beta}$. The fitted coefficient $\beta$ consistently remains below two for all temperatures studied, indicating that the particles do not undergo ballistic motion at short times.
This deviation from ballistic motion arises from initial particle rearrangements within the interacting potential range, including occasional jumps between lattice points.
After the quasi-ballistic regime, the MSD enters a plateau, indicating transient dynamical arrest as  particles become confined within their local environments. In crystalline solids, such plateaus  reflect vibrational motion around lattice sites, while  in glassy and quasicrystalline systems, they result from  local structural constraints and energy barriers that hinder diffusion. As temperature increases, these  barriers become easier to overcome, leading to shorter plateaus and  indicating that thermal fluctuations facilitate cage escape. At longer times, particles regain mobility through collective rearrangements or by hopping between energetically equivalent positions, consistent with phason-like dynamics in quasicrystals.

This behavior is evident in Fig.~\ref{fig:msd-diffusion}a, which shows that at longer times, increasing the temperature enables particles to  overcome local energy barriers, allowing the MSD to transition into a diffusive regime characterized by the scaling \(\left\langle r^{2}(t) \right\rangle \sim t\) with a temperature-dependent slope. 
Importantly, within the temperature range studied  (\(k_{B} T / \epsilon \leq 0.3\)), this long-time diffusion occurs while  global quasicrystalline order remains intact, as confirmed by  the RDF and diffraction patterns.
At lower temperatures, particles remain trapped near  their equilibrium positions, and the structure remains both dynamically and structurally stable.

A clearer insight into this mechanism is provided by the self-diffusion coefficient, defined as $D= \lim_{t \rightarrow  \infty} (d \left\langle r^{2}(t) \right\rangle/dt)/6 $. As shown in Figure~\ref{fig:msd-diffusion}b, the diffusion coefficient exhibits an Arrhenius-like dependence on temperature
\begin{equation}
    D \tau/\sigma^2 = D_{0} \exp{\left( - \Delta E / k_{B} T \right)} \, ,
    \label{eq:arrhenius}
\end{equation} 
where \(D_0\) is a prefactor and \(\Delta E\) is the activation energy. In this system, we identify two distinct dynamical regimes characterized by different activation energies: \(\Delta E = 1.12(1) ~k_BT\) in the low-temperature regime and \(\Delta E = 3.92(2)~k_BT\) at  high temperatures. 

At low temperatures, particle motion is dominated by phason-assisted jumps between energetically equivalent sites, consistent with the aperiodic order of the quasicrystal~\cite{kaluginMechanismSelfDiffusionQuasiCrystals1993,engelDynamicsParticleFlips2010}. As  temperature increases, the dynamics becomes more complex, involving a combination of individual hopping events, defect-mediated processes, and cage-breaking motions, reflecting an interplay of multiple diffusion mechanisms.

\begin{figure}
    \includegraphics[height=5in]{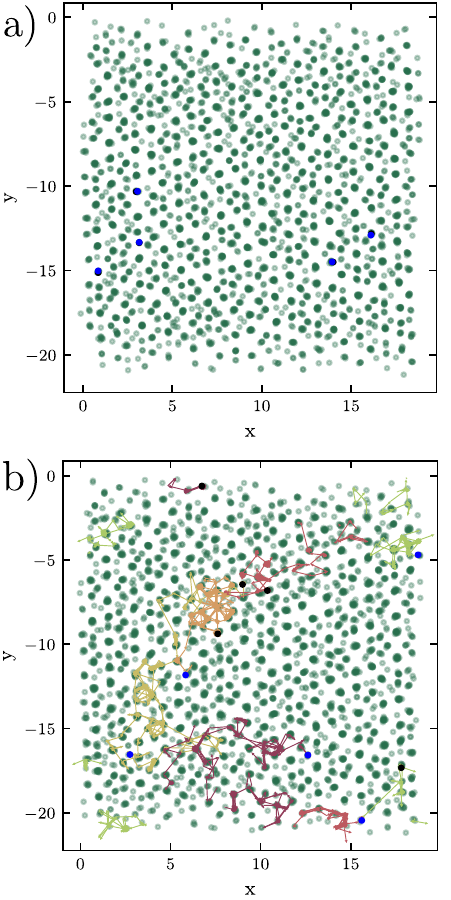}
    \caption[traj]{Single-particle trajectories in the icosahedral quasicrystal, observed along the five-fold symmetry axis. The blue dot denotes  the initial position, and  black represents the final position after a  time window  \([2 \times 10^{7} \taumd, 2.5 \times 10^{7} \taumd]\), at a constant pressure of \(P \sigma^{3} / \epsilon = 0.1\).
    \begin{enumerate*}[label=\alph*)]
        \item At temperature \(k_{B} T / \epsilon = 0.15\), particles remain close to their starting positions, with motion restricted to  vibrations and occasional small jumps. This state point corresponds to the low-temperature regime with suppressed diffusion  (lower curve in Fig.~\ref{fig:msd-diffusion}b).\label{fig:low-temp}
        \item At temperature \(k_{B} T / \epsilon = 0.3\),  particles exhibit larger displacements along symmetry-related particle positions, following the quasicrystalline order while enabling longer-range motion.   This statepoint corresponds to the higher-temperature diffusion regime (upper  curve in Fig.~\ref{fig:msd-diffusion}b).\label{fig:high-temp}
    \end{enumerate*}
    }
    \label{fig:mechanisms}
\end{figure}

Both diffusion mechanisms are observed in our simulations,  depending on temperature. We show in Fig.~\ref{fig:mechanisms}, the motion of selected particles at two temperatures viewed along the five-fold symmetry axis of the IQC.
At \(k_{B} T / \epsilon = 0.15\) and \(P \sigma^{3} / \epsilon = 0.1\), shown in Fig.~\ref{fig:mechanisms}a, particles remain close to their initial positions, exhibiting only vibrational motion.
In contrast, at a higher temperature of \(k_{B} T / \epsilon = 0.3\), particles acquire sufficient energy to  overcome local barriers, leading to collective motion and diffusion. This is evident  in Fig.~\ref{fig:mechanisms}b, where single-particle trajectories overlap, indicating shared  paths and correlated motion.
These findings are consistent with  observations  in two-dimensional quasicrystals~\cite{engelDynamicsParticleFlips2010,zhao2025atomistic}, suggesting that such collective dynamics is a general feature of quasicrystals and can occur in both two or three dimensions.

\subsection{\label{sec:heterogeneity}Dynamic heterogeneity}
\begin{figure}
    \includegraphics[width=\linewidth]{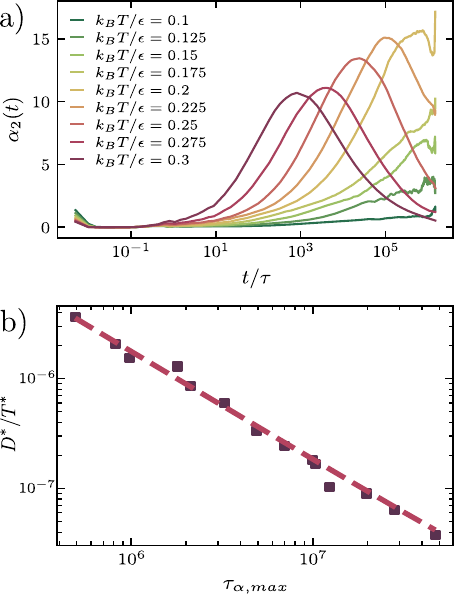}
    \caption[ngauss]{\begin{enumerate*}[label=\alph*)]
       \item Non-Gaussian parameter $\alpha_2(t)$ as a function of time $t$ of an icosahedral quasicrystal, revealing increasing dynamic heterogeneity with increasing temperature,  reaching a maximum at  \(k_{B} T / \epsilon = 0.2\),  after which it gradually decreases.
       \item The reduced long-time self-diffusion coefficient \(D^{*} / T^{*}\), with \(D^{*} = D \tau / \sigma^{2}\) and \(T^{*} = k_{B}T / \epsilon\), as a function of the time at which the maximum value of the non-Gaussian parameter is observed, \(\tau_{\alpha,\text{max}}\). The line is a non-linear fit to the data. \label{fig:scaling}
       \end{enumerate*}}
    \label{fig:non-gauss-power-law}
\end{figure}

The global dynamics, as captured by the self-diffusion, emerge from a complex interplay of multiple  mechanisms. Furthermore, the diversity of local structural environments within the quasicrystal  leads to variations in particle mobility, giving rise to dynamic heterogeneities across the system, phenomena already observed in two-dimensional systems.~\cite{zhao2025atomistic} To quantify these  dynamic heterogeneities, we calculate the non-Gaussian parameter \(\alpha_{2}(t)\), defined as 
\begin{equation}
    \alpha_{2}(t) = \frac{3 \left\langle r^{4}(t) \right\rangle}{5 {\left\langle r^{2}(t) \right\rangle}^{2}} - 1 \, .
\end{equation}
The time evolution of \(\alpha_{2}(t)\) is shown in Fig.~\ref{fig:non-gauss-power-law} for different temperatures. The behavior of this parameter is similar  to that observed in glass-forming systems~\cite{kobTestingModecouplingTheory1995c,adhikari2021spatial} and crystalline systems with vacancy-mediated dynamics~\cite{meerDynamicalHeterogeneitiesDefects2015}.
At short times, $\alpha_{2}$ remains negligible across all temperatures, indicating that the dynamics is homogeneous in this regime.
At longer times, however, we observe the emergence of dynamical heterogeneity, indicated by $\alpha_{2}>0$, with its magnitude depending on  temperature. At the lowest temperatures,  particles remain near  their lattice points, resulting  in  relatively  homogeneous dynamics. 
As the temperature increases, particles gain sufficient energy to overcome local energy  barriers, triggering collective rearrangements. This leads to clear spatial dynamical heterogeneity, with   regions of increased mobility interspersed with regions where  particles remain largely immobile.

The timescale at which $\alpha_2(t)$  peaks, \(\tau_{\alpha,\text{max}}\), indicates   the time at which the system exhibits the largest dynamical  heterogeneity. Since the system is in equilibrium,  \(\alpha_{2}(t)\) returns  to zero after this time as  homogeneity  is restored and particles  diffuse uniformly.
Both the characteristic time \(\tau_{\alpha,\text{max}}\) and the peak value $\alpha,\text{max}$ decrease with increasing temperature, reflecting a transition toward more homogeneous dynamics as the system approaches the fluid phase. Notably, in Fig.~\ref{fig:non-gauss-power-law}\ref{fig:scaling} we observe a linear relationship between the diffusion coefficient and the inverse of $\alpha_2^{\text{max}}$, further linking the extent of dynamical heterogeneity to particle mobility in the system.
This trend is commonly observed in glass-forming materials~\cite{marin2020tetrahedrality,kobTestingModecouplingTheory1995c,adhikari2021spatial}, suggesting that the dynamics of icosahedral quasicrystals at low temperatures share similarities with those of glass-forming liquids.  Similar behavior has also been  reported in  two-dimensional dodecagonal 
quasicrystals~\cite{zhao2025atomistic}.

\section{\label{sec:structure-dynamics}Linking Quasicrystalline Structure to Dynamical Behavior}
\begin{figure}
    \includegraphics[width=0.9\linewidth]{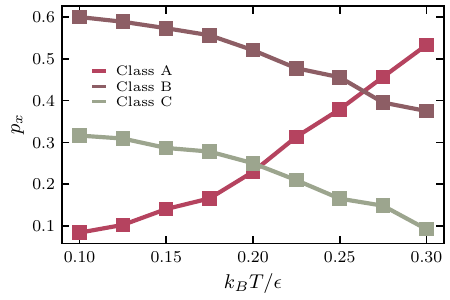}
    \caption{Fraction of particles belonging to a specific class \(p_{x}\), where $x$ denotes either class A, B, or C, as a function of reduced temperature within the regime where the quasicrystal is found to be stable.}
    \label{fig:class-temperature}
\end{figure}

To explicitly connect our findings on structure and dynamics, we compute the order parameter \(p_{x}\) for equilibrated samples of the icosahedral quasicrystal at the same temperatures used in our dynamical analyses. Using our UML framework, we determine \(p_{x}\) for each state point and plot it as a function of reduced temperature in Fig.~\ref{fig:class-temperature}.

At the lowest temperature, \(k_{B} T / \epsilon = 0.1\), we observe that class~A, associated with configurations exhibiting less quasicrystalline order and typically appearing in the liquid phase before the IQC formation, has the lowest order parameter value, \(p_{x} \lesssim 0.1\). In contrast, classes~B and~C show significantly higher values of \(p_{x}\), with class~B reaching the highest. This indicates a pronounced degree of structural order, consistent with the presence of pentagonal, icosahedral and dodecahedral motifs.
These observations reveal a clear correlation: high structural order within the IQC coincides with suppressed long-range diffusion, where defects are rare and particles remain localized near their equilibrium positions. This behavior is supported by observations at higher temperatures.

At \(k_{B} T / \epsilon = 0.2\), there is a noticeable increase in class~A, with \(p_{x} \gtrsim 0.2\), up from its nearly negligible value at \(k_{B} T / \epsilon = 0.1\). Accordingly, at this temperature the  the system enters the collective rearrangement regime, with increased diffusion.  This suggests that class~A plays a key role in the onset of diffusive dynamics. Since this class also appears in the liquid just before IQC formation, it is reasonable to associate it with collective particle motion. If this hypothesis holds, it implies that collective dynamics, encoded in class~A, combined with the phason-like motion characteristic of classes~B and~C, facilitates particle self-diffusion within the IQC.

Finally, at \(k_{B} T / \epsilon = 0.3\), class~A reaches its highest value, \(p_{x} \gtrsim 0.55\), while class~C decreases significantly to \(p_{x} \lesssim 0.1\). Notably, class~B remains present with \(p_{x} \approx 0.4\), indicating that a considerable degree of pentagonal ordering  persists even at this elevated temperature. As shown by the non-Gaussian parameter \(\alpha_{2}(t)\) in Fig.~\ref{fig:non-gauss-power-law}, dynamical heterogeneity remains pronounced. These observations suggest that the dominant mode of diffusion within the IQC is collective in nature: as  the icosahedral and pentagonal ordering diminishes, particles become more mobile, enabling enhanced structural rearrangements across the system. This highlights a strong interplay between temperature, quasicrystalline structure, and particle dynamics across the IQC stability regime. In particular, increasing temperature leads to a shift from localized, phason-like fluctuations toward collective particle rearrangements that facilitate diffusion while partially preserving quasicrystalline order.

\section{\label{sec:conclusions}Conclusions}

In this work, we have investigated the structure and dynamics of a one-component model system of isotropic particles interacting with an oscillating pair potential that stabilizes an icosahedral quasicrystal (IQC). Our results reveal  that the IQC remains stable within a narrow region of the phase diagram, particularly over a well-defined range of pressure and temperature.

The equation of state along a quasicrystalline isotherm reveals the delicate stability of the IQC, constrained to a narrow range of  pressure-density conditions. Upon compression, the IQC transitions into an amorphous solid.
A similar structural progression is observed along a liquid isotherm. Analysis of the radial distribution function reveals the unique structural motifs of the IQC, characterized by a distinct second shell of neighbors arranged in icosahedral and dodecahedral patterns. These arrangements are reminiscent of those found in certain glass-forming systems~\cite{coslovichDimensionalityReductionLocal2022b}, highlighting the structural parallels between  IQCs and amorphous solids.

To deepen our understanding of these structural relationships, we employed averaged bond-orientational order parameters (BOPs) in combination with an unsupervised machine learning (UML) framework for dimensionality reduction and clustering. This approach enables us to classify the various phases--fluid, amorphous solid, FCC crystal, and IQC--using  solely  structural data. Remarkably, the application of this UML framework  to the IQC reveals three distinct local particle environments, providing valuable insight into the kinetic pathways underlying IQC formation.
We find that during the formation of the IQC from the fluid certain low-coordination geometric motifs emerge in the fluid phase prior to the transformation. These motifs then act as  building blocks for the high-coordination structures, such as icosahedra and dodecahedra, that characterize  the quasicrystal.

Our findings further demonstrate that the IQC exhibits dynamic heterogeneities, as captured by the non-Gaussian parameter.
We identify two distinct dynamical regimes. At intermediate temperatures, the dynamics is predominantly governed by activated processes, consistent with the Arrhenius behavior observed in the long-time diffusion. In contrast, at low temperatures, the dynamics is primarily governed by phason-like excitations.
Single-particle trajectories along the five-fold symmetry axis at different temperatures confirm this crossover in dynamical behavior.

To connect structure and dynamics, we analyzed how the relative distributions of particles belonging to UML-identified structural classes depend on temperature. We found that high structural order (e.g., classes B and C, associated with pentagonal and icosahedral environments) correlates with suppressed self-diffusion. Conversely, lower structural order (class A), which is prevalent near the onset of quasicrystal formation and at elevated temperatures, is associated with enhanced collective motion. This direct correspondence between local structure and dynamical behavior offers a microscopic interpretation of how phason-like and activated dynamics emerge and compete within IQCs.
In addition, the UML framework established here provides a versatile tool for  identifying local structures in quasicrystals with different symmetries, making it valuable for studying the transformation mechanisms in other quasicrystalline systems.

Looking ahead, integrating our structure-dynamics framework with high-throughput machine learning models could enable the predictive design of novel quasicrystals, building on recent advances in ML-guided discovery of stable quasicrystalline phases. Moreover, elucidating the connection between local structure and dynamics may inform the engineering of quasicrystalline materials in applications such as phononic or photonic crystals, where controlling defect-tolerant and diffusion-limited quasi-crystalline domains is essential. Finally, extending this framework beyond metallic systems to soft-matter or photonic quasicrystals could reveal universal structure-mobility relationships across a broad class of quasicrystalline materials.

Finally, this UML framework can also be applied to investigate the structural order of supraparticles---colloidal assemblies formed  under spherical confinement that exhibit icosahedral point group symmetry~\cite{denijsEntropydrivenFormationLarge2015,wangBinaryIcosahedralClusters2021}. Extending these methodologies could accelerate the discovery of IQCs in experimentally realizable colloidal or atomic systems, thereby bridging the gap between theoretical predictions and experimental realizations.

\begin{acknowledgments}
The authors acknowledge funding from the European Research Council (ERC) under the European Union's Horizon 2020 research and innovation program (Grant agreement No. ERC-2019-ADG 884902, SoftML).
\end{acknowledgments}
\section*{Author Contributions}

E.A.B.-M.: Conceptualization, Methodology, Software, Validation, Formal Analysis, Investigation, Data Curation, Visualization, Writing - Original Draft, Writing - Review and Editing. S.M.-A.: Conceptualization, Methodology, Software, Validation, Investigation, Writing - Original Draft, Writing - Review and Editing. M.D.: Conceptualization, Resources, Supervision, Project administration, Funding acquisition, Writing - Original Draft, Writing - Review and Editing.

\section*{Conflict of interest}
The authors have no conflicts to disclose.

\section*{Data Availability}
The data that support the findings of this study are available from the corresponding author upon reasonable request.

\appendix

\section{\label{ap:nucleation}Cooling and compression protocol}
\begin{figure}
    \includegraphics[width=\linewidth]{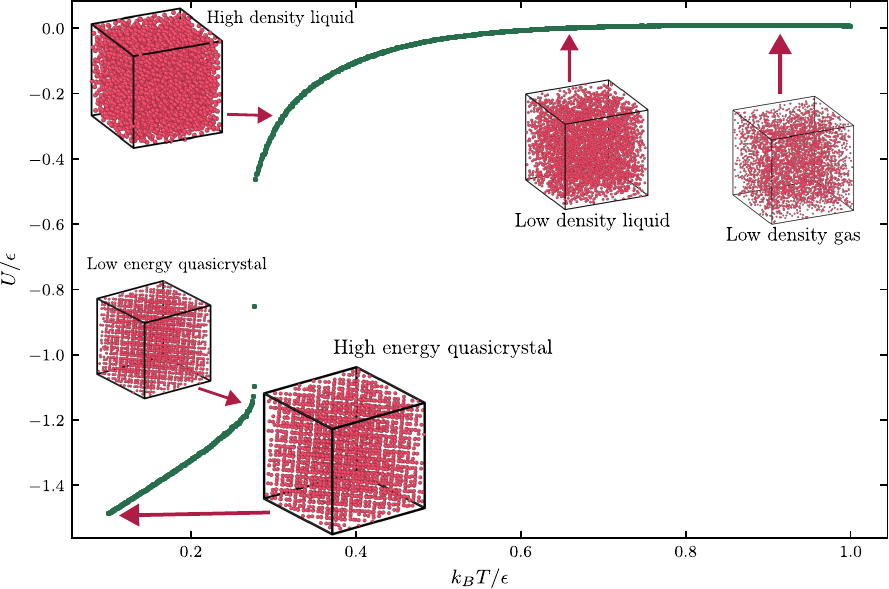}
    \caption{Formation of the icosahedral quasicrystal by starting at a low-density liquid and compressing it as well as cooling it until the quasicrystal spontaneously forms. The simulation continues in order to obtain a lower-energy system and reaching the target temperature.}
    \label{fig:energy-cooling}
\end{figure}

A schematic representation of the cooling protocol is provided in Fig.~\ref{fig:energy-cooling}, along with snapshots illustrating the pathway leading to the formation of quasicrystals.
The reduced energy per particle, $U/\epsilon$, was monitored throughout the cooling process, revealing the evolution of energy to lower values consistent with the stabilization of the quasicrystal phase, as well as a sharp transition from the liquid to the quasicrystal.

The system was initialized in a high-temperature, low-pressure liquid state characterized by \(k_{B}T / \epsilon = 1\) and \(P \sigma^{3} / \epsilon = 0.01\).
From this initial condition, the system was subjected to a combined cooling and compression protocol, during which it traversed a sequence of liquid states, displayed in Fig.~\ref{fig:energy-cooling}.
Upon gradual cooling and compression, a quasicrystalline phase spontaneously emerged from the liquid. Based on the energetic signatures shown in Fig.~\ref{fig:energy-cooling}, this phase transition  occured at approximately \(k_{B}T / \epsilon \approx 0.35\). Further reduction of temperature and increase in pressure  led to the formation of a low-energy icosahedral quasicrystal (IQC) at the final state point of \(k_{B}T / \epsilon = 0.1\) and \(\beta P \sigma^{3} = 0.1\). This structure exhibits the characteristic two-, three-, and five-fold rotational symmetries of icosahedral quasicrystals.

\begin{figure}
    \includegraphics[width=\linewidth]{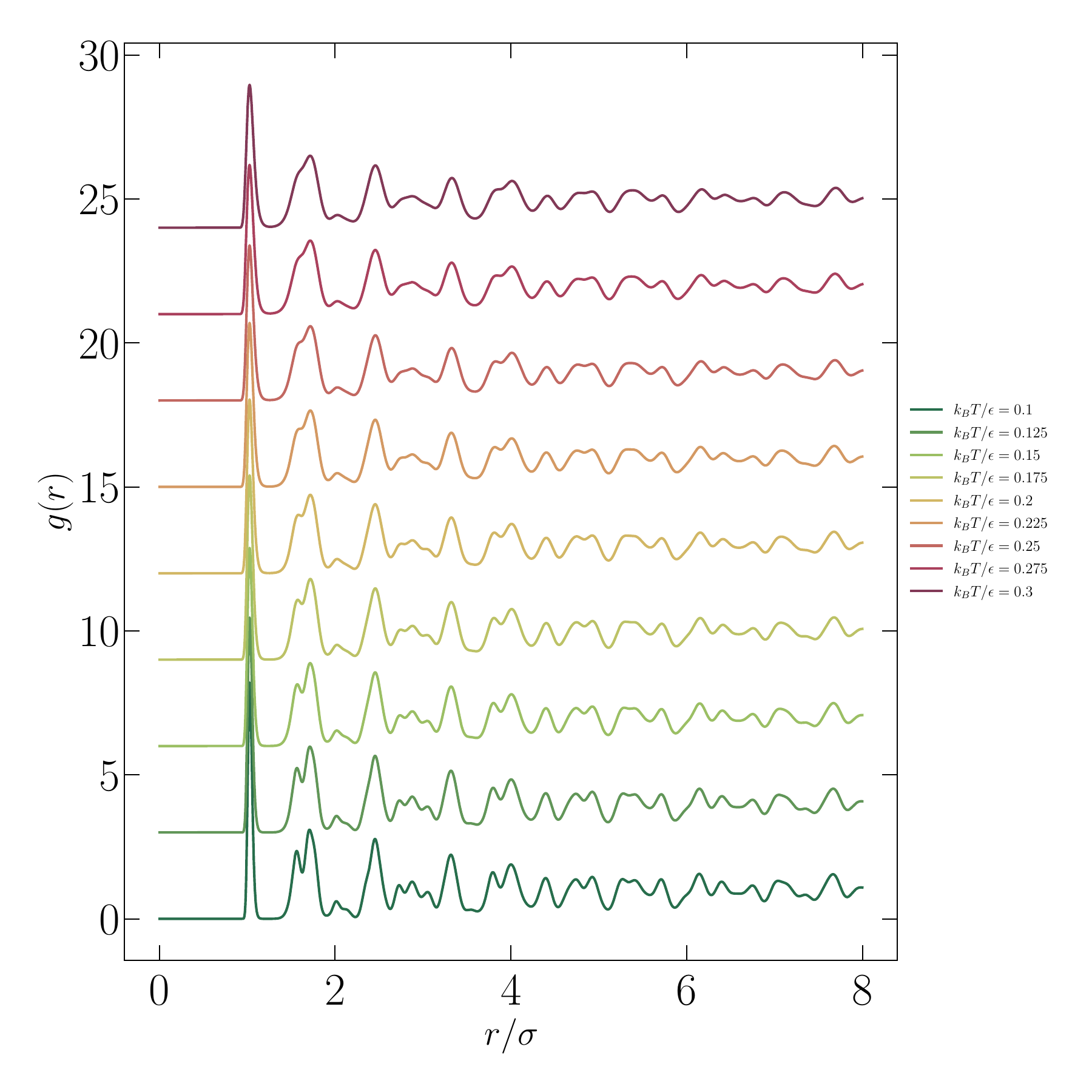}
    \caption{Radial distribution function of an icosahedral quasicrystal at different temperatures.}
    \label{fig:qc-rdf}
\end{figure}

Within the temperature range \(k_{B}T / \epsilon \in [0.1, 0.3]\), the quasicrystalline structure remains stable, as evidenced by the radial distribution function shown in Fig.~\ref{fig:qc-rdf}. The presence of long-range orientational order is demonstrated by a series of pronounced peaks in the radial distribution function, corresponding to successive coordination shells. These features are indicative of the quasi-periodic tiling that underpins the quasicrystalline arrangement.

In the short-range regime, the two secondary peaks following the principal peak at \(r/\sigma = 1\) align with the energy wells in the pairwise interaction potential, confirming local ordering consistent with icosahedral and dodecahedral motifs. Importantly, the intensity of these peaks diminishes with increasing temperature, suggesting a progressive loss of local structural order.
At larger distances, persistent fluctuations reflect the absence of long-range translational symmetry, a defining feature of quasicrystalline phases.

\section{\label{appendix:umap} UMAP and Gaussian Mixture Model hyperparameters}
To cluster and classify different phases, as well as to distinguish different local environments in the quasicrystal, we employ an unsupervised machine learning (UML) framework based on the bond orientational order parameters introduced in Sec.~\ref{sec:order}. Specifically, we use the Uniform Manifold Approximation and Projection (UMAP) technique~\cite{mcinnes2020umapuniformmanifoldapproximation}.
This approach  projects high-dimensional data into a low-dimensional space, facilitating the classification of phases with distinct structural features.

We first project the data onto a two-dimensional space using UMAP. We use the official implementation of Ref.~\citenum{mcinnes2020umapuniformmanifoldapproximation}, where we set the hyperparameters \verb|n_neighbors = 100| and \verb|min_dist = 0|.
This choice yields compact, well-separated clusters suitable for unsupervised classification.
The hyperparameter \verb|n_neighbors| determines the size of the local neighborhood (in terms of number of neighboring sample points) used for manifold approximation. Larger values emphasize the global structure of the underlying manifold, producing larger clusters while reducing noise from  fine-grained local variations.

In contrast, the hyperparameter \verb|min_dist| determines the effective minimum distance between embedded points, with smaller values yielding more compact and clustered embeddings.
We also tested smaller values of \verb|n_neighbors| and observed no significant changes in the clustering outcomes.
However, the choice of \verb|min_dist| was found to have a stronger influence: increasing it leads to less dense projections, which can reduce the resolution of the resulting clusters.

We then apply a Gaussian mixture model (GMM)~\cite{hastie2017elements} to cluster the projected data. The GMM  assumes that the data points are generated from a mixture of a finite number of Gaussian distributions with unknown parameters~\cite{hastie2017elements}. We use the implementation of \verb|scikit-learn|~\cite{scikit-learn} specifying a full covariance matrix, so that each component can adopt an arbitrary shape.

Given the four phases present in our input dataset, we set the number of  components in the mixture model to four.
The GMM requires an initial  estimate of the mean and covariance of the clusters, a process commonly referred to as "seeding"~\cite{hastie2017elements}. For our purpose, we employ the seeding approach provided by the \(k\)-means++ algorithm~\cite{ilprints778} which provides a straightforward initialization via   random seeding and enables the GMM to converge with fewer function calls.

\section*{Data Availability}
The data that support the findings of this study are available from the corresponding author upon reasonable request.

\bibliography{bibliography}

\end{document}